\documentclass[10pt]{article}
\usepackage{amsmath}
\usepackage{amssymb}
\usepackage{graphicx}
\usepackage{cite}

\usepackage[labelfont=bf,labelsep=period,justification=raggedright]{caption}
\makeatletter
\renewcommand{\@biblabel}[1]{\quad#1.}

\makeatother

\date{}
\begin{document}

\begin{flushleft}
{\Large
\textbf{A Human Development Framework for CO$_2$ Reductions}
}

Lu\'is Costa$^{\ast}$, 
Diego Rybski, and 
J\"urgen P. Kropp
\\
Potsdam Institute for Climate Impact Research, Potsdam, Germany

$^{\ast}$ E-mail: carvalho@pik-potsdam.de
\end{flushleft}

\section*{Abstract}

Although developing countries are called to participate in CO$_2$ emission reduction efforts to avoid dangerous climate change, the implications of 
proposed reduction schemes in human development standards of developing countries remain a matter of debate. We show the existence of a positive and time-dependent correlation between the Human Development Index (HDI) and per capita CO$_2$ emissions from fossil fuel combustion. Employing this empirical relation, extrapolating the HDI, and using three population scenarios, the cumulative CO$_2$ emissions necessary for developing countries to achieve particular HDI thresholds are assessed following a Development As Usual approach (DAU).

If current demographic and development trends are maintained, we estimate that
by 2050 around 85\% of the world's population will live in countries with high
HDI (above 0.8). In particular, 300\,Gt of cumulative CO$_2$ emissions between 2000 and 2050 are estimated to be necessary for the development of 104 developing countries in the year~2000. This value represents between 20\,\%~to~30\,\% of previously calculated CO$_2$ budgets limiting global warming to $2\,^{\circ}\mathrm{C}$.

These constraints and results are incorporated into a CO$_2$ reduction
framework involving four domains of climate action for individual countries. The
framework reserves a fair emission path for developing countries to proceed with
their development by indexing country-dependent reduction rates
proportional to the HDI in order to preserve the $2\,^{\circ}\mathrm{C}$
target after a particular development threshold is reached. For example, in each time step of five years, countries with an HDI of $0.85$ would need to reduce their per capita emissions by approx. $17$\,\% and countries with an HDI of $0.9$ by $33$\,\%. Under this approach, global cumulative emissions by 2050 are estimated to range from 850 up to 1100\,Gt of CO$_2$. These values are within the uncertainty range of emissions to limit global temperatures to $2\,^{\circ}\mathrm{C}$.

\section*{Introduction}

Consensus emerging in favor of low CO$_2$ stabilization targets requires the participation of developing countries in the efforts to reduce global green-house emissions \cite{Elzen2008}. For example, it has been claimed that in order to keep global temperatures below a $2\,^{\circ}\mathrm{C}$ increase, developing countries should attain more than 20\,\% CO$_2$ reductions below business-as-usual levels by the year 2020 \cite{Rogelj2010}. The potential implications of such reductions on development standards remain unclear \cite{Cocklin2007} as developing countries are expected to extensively rely on fossil energy to fuel their current development needs \cite{Bolin2001}. In addition to potential development implications, a fair allocation of responsibility regarding CO$_2$ emissions reduction between developed and developing countries remains a controversial topic \cite{Broecker2007,Chakravarty2010}. How to account for the responsibility of developed countries regarding historical CO$_2$ emissions \cite{WBGU2006} and to what extent technological and political inertia impose limits to the range of strategies envisioning the implementation of reduction schemes \cite{IPCC2001} are questions that remain largely unanswered. Developing countries have expressed their concerns on the points raised, questioning if development goals can -- or cannot -- be met under current technological and population trends \cite{Pan2005}. 

In order to tackle above mentioned challenges, the CO$_2$ allocation and reduction approach here outlined contrasts from existing ones \cite{WBGU2006,Broecker2007,Baer2008} by relying on the Human Development Index (HDI) \cite{UNDP2008} as a summary measure reflecting the achievement of a country in three basic dimensions of human development: a long healthy life, access to knowledge, and decent living standards. These dimensions are assessed based on the following indicators: life expectancy at birth, literacy rate of adults, gross enrollment rate, and gross domestic product per capita at purchasing power parity \cite{UNDP2008}. Despite some criticism -- for example treating income, health, and education as substitutes \cite{Neumayer2001} --  the HDI has been consistently used by the United Nations Development Programme (UNDP) as a reference metric to compare social and economic development within and between countries across time. Furthermore, the HDI has been reported to play an important role in raising the political profile of general health and educational policies \cite{Atkinson1997}, to be an indicator of a country's exposure to climate-related extremes \cite{Patt2010} and its dimensions determinants of vulnerability and adaptive capacity at national level \cite{Brooks2005}.

\begin{figure}
\includegraphics[width=0.6\textwidth]{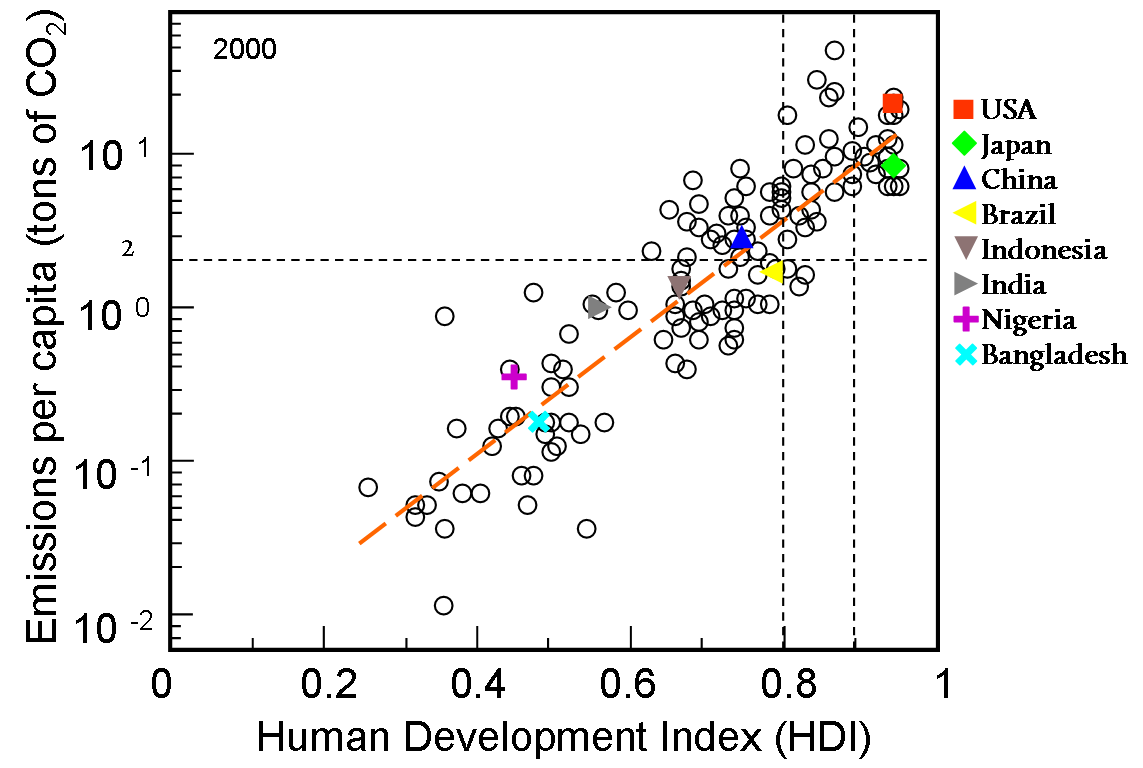}
\caption{
{\bf Correlations between~HDI and~CO$_2$ per capita emissions in the year 2000.}
The dashed line represents a least squares fit through all values. The
coefficient of determination is $R^2\simeq 0.81$ and the correlation coefficient
is $\rho\simeq 0.90$. For some countries the values are shown explicitly.
Vertical lines represent the HDI values of 0.8 and 0.9 representative of high
and very high development standards respectively as expressed in the United
Nations Development Report 2009 \cite{Bongaarts2010}. The horizontal line shows
the 2~tons per capita CO$_2$ emissions target to limit global warming at
$2\,^{\circ}\mathrm{C}$ by 2050 \cite{WBGU2006}.
}
\label{fig:correlations}
\end{figure}

In Figure~\ref{fig:correlations} per capita emissions are plotted against the
corresponding HDI for countries with available data in the year 2000. We find that the per capita CO$_2$ emissions from
fossil fuel burning are exponentially correlated with human 
development -- highlighting the often disregarded 
social-dimension of emissions reductions. 
For example, the development strategy
targeting high growth in domestic product by relying on low-cost, low-efficiency
technology, contributed for the poverty rate in China to drop from 53\% in 1981
to 8\% in 2001 \cite{Montalvo2010}. Although this "fossil" path of development
is highly incompatible with future climate targets, climate policies cannot
neglect the potential societal implications of CO$_2$ reductions, especially
during the first stages of human development in a country. The magnitude of the
challenges ahead become clear once the per capita CO$_2$ emissions guard rail of 2 tons for avoiding dangerous climate change \cite{Schellnhuber2006} and the HDI thresholds of 0.8 and 0.9 (characteristic of a developed world) are displayed. A fair distribution of CO$_2$ emissions under current technological constraints should allow the convergence of developing countries towards 0.8 or 0.9 HDI scores and, at the same time, keep global CO$_2$ emissions below the available budgets limiting anthropogenic climate change.

\section*{Methods}
\subsection*{Extrapolating the Human Development Index}

\begin{figure}
\includegraphics[width=0.5\textwidth]{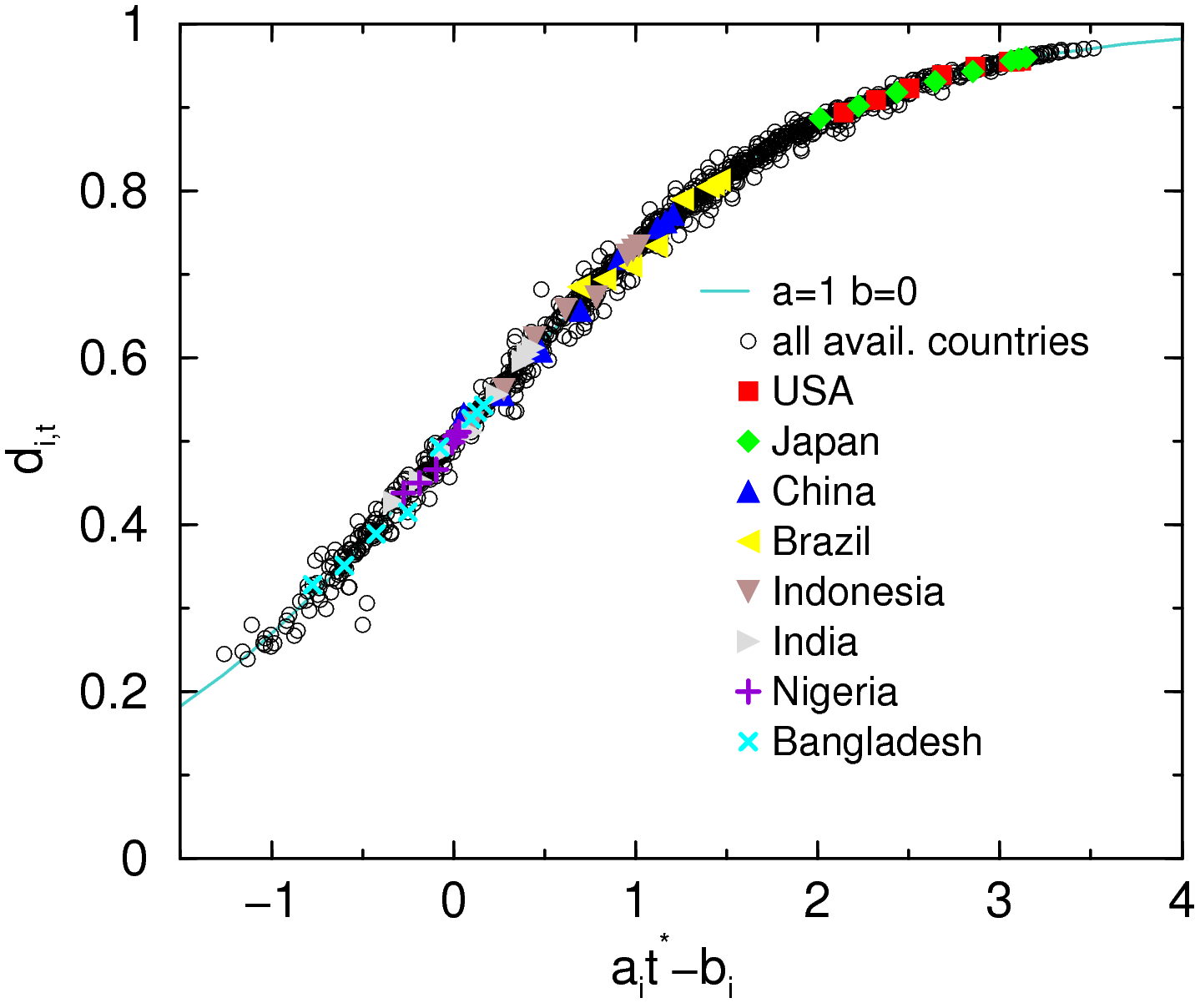}
\caption{
{\bf Collapse of the HDI values based on logistic regression according to 
Eq.~(\ref{eq:hdilogreg}).}
HDI values are plotted for each country by using a transformed time
$t^*=\frac{t+b_i}{a_i}$ so that HDI values of all countries (open circles) fall
within their spreading on the curve which is used to fit the data. The filled
symbols intend to highlight the same examples as in Fig.~\ref{fig:correlations}.
The solid line corresponds to the function $d_{t} = \frac{1}{1+{\rm e}^{-t}}$.
}
\label{fig:hdicollaps}
\end{figure}

Our approach starts by investigating the evolution of future human development standards. We assume that the HDI, $d_{i,t}$, of a country, $i$, evolves in time, $t$, following a logistic regression \cite{HosmerL2000}. This choice is supported by the fact that the HDI is bounded to $0\le d_{i,t}\le 1$ and that countries with high HDI evolve slowly in time. Further, this asymptotic behavior suggests the existence of smooth transitions in development. The logistic regression fulfills these requirements.
Therefore, we fit for each country separately
\begin{equation}
\label{eq:hdilogreg}
\tilde{d}_{i,t} = \frac{1}{1+{\rm e}^{-a_it+b_i}}
\end{equation}
to the available data (obtaining the parameters~$a_i$ and~$b_i$). Since the regression involve only two parameters, three measures of HDI would suffice to over-determine the system. We have chosen to use countries for which we have at least four~values of HDI in order to obtain more robust results. This lead to regressions for $147$~countries out of $173$ in our data set. Basically, $a_i$ quantifies how fast a country develops and $b_i$ represents a delay. In Figure~\ref{fig:hdicollaps} we display the collapse (see e.g. \cite{MalmgrenSCA2009}) of the past HDI as obtained from the logistic regressions illustrating how countries have been developing in the scope of this approach. The HDI values of each country are plotted using a transformed time $t^*=\frac{t+b_i}{a_i}$ so that values of all countries (open circles) fall within their spreading on the curve which is used to fit the data. The filled symbols highlight the same countries as in Figure~\ref{fig:correlations}. The solid line corresponds to the function $d_{t} = \frac{1}{1+{\rm e}^{-t}}$. Based on the obtained parameters,~$a_i$ and~$b_i$, we estimate the future HDI of each country until 2050 assuming similar development trajectories as in the past. 

\subsection*{Projecting per capita emissions}

In the following section we provide the main assumptions used to extrapolate per capita emissions of CO$_2$ from fossil fuel burning (see also section III.B in Text S1). We choose not to include emissions from other greenhouse gases since they were found not to be strongly correlated with personal consumption and national carbon intensities \cite{Chakravarty2010}. CO$_2$ emissions from land-use were disregarded due to the high uncertainty of historical data \cite{Rhemtulla2008}.

The correlations between HDI and CO$_2$ emissions per capita, $e^{(\rm c)}_{i,t}$, were assessed for all years (1980-2006), see example of Figure~\ref{fig:correlations}. We apply the exponential regression
\begin{equation}
\label{eq:co2hdiexp}
\hat{e}^{(\rm c)}_{i,t} = {\rm e}^{h_{t} d_{i,t} + g_{t}}
\end{equation}
to the data by linear regression \cite{MasonGH1989} through
$\ln e^{(\rm c)}_{i,t}$ versus $d_{i,t}$ for fixed years~$t$ and obtain the parameters $h_{t}$ and $g_{t}$. 
At a global level, correlation coefficients varied between a minimum of $0.89$ in 2005 and a maximum of $0.91$ in 2006. The individual components of HDI were found to be as well correlated with per capita emissions, in the following decreasing order of correlation coefficient: GDP, education, and life expectancy, see Figure~S2 and Table~S2 in Text S1.

We take advantage of these correlations and assume that the system is ergodic, i.e. that the process over time and over the statistical ensemble is the same. In other words, we assume that these correlations also hold for each country individually and apply the exponential regression
\begin{equation}
\label{eq:co2hdiexpc}
\tilde{e}^{(\rm c)}_{i,t} = {\rm e}^{h_{i} d_{i,t}+g_{i}}
\enspace 
\end{equation}
obtaining the parameters $h_{i}$ and $g_{i}$, which are now country dependent. Based on the estimated parameters the CO$_2$ per capita emissions are extrapolated country wise. We additionally tested two population-weighing methods when fitting per capita emissions versus the HDI (see section~III.E and Figure~S7 in the Text S1).  

For $52$~countries out of $173$ data was found to be insufficient to perform the regressions Eq.~(\ref{eq:hdilogreg}) or Eq.~(\ref{eq:co2hdiexpc}). This is, they comprise less than the minimum number of data points required to fit the HDI versus time or CO$_2$ emissions per capita versus HDI. In the Text S1 (see section~III.C and Figures~S3 and~S4), we find that changes of $d_{i,t}$ and $e^{(\rm c)}_{i,t}$ are correlated among the countries. Thus, in the $\ln e_{i,t}^{\rm (c)}$-$d_{i,t}$-plane, we let countries with a lack of data evolve in a similar way as those in their vicinity. 

\begin{figure}
\includegraphics[width=0.85\textwidth]{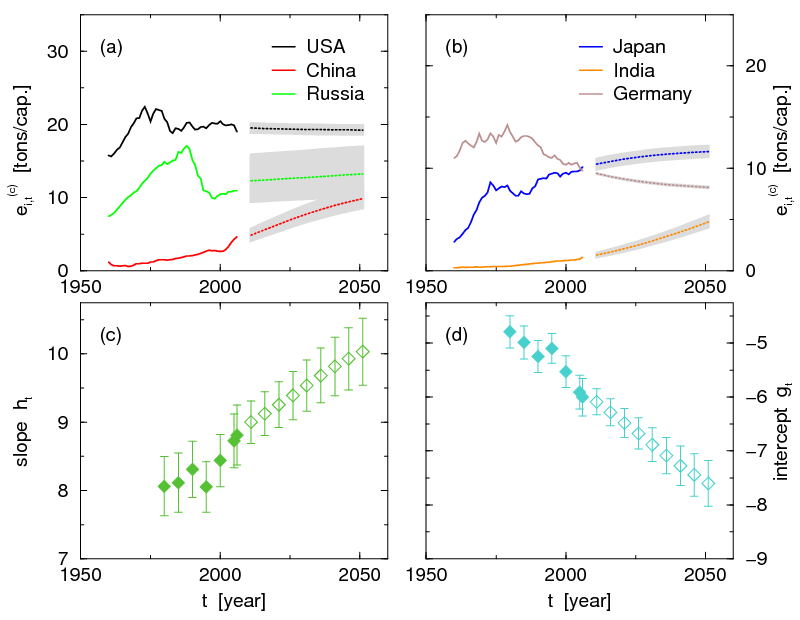}
\caption{
{\bf Examples of extrapolated CO$_2$ per capita emissions.} 
Panels {\bf a} and {\bf b} show the extrapolated values of CO$_2$ emissions per
capita for 6 countries following a DAU approach. The gray uncertainty range is
obtained by including the statistical errors of the regressions (one SD each).
Panels {\bf  c} and {\bf  d} represent the slope and intercept values for the
ensemble regressions of HDI versus CO$_2$ per capita for observed (filled
symbols) and projected (open symbols) data. The error bars are given by the
standard errors.
}
\label{fig:examples}
\end{figure}

In Figure~\ref{fig:examples} the panels~(a) and~(b) show examples of extrapolated CO$_2$ emissions per capita for six countries according to the described methodology (more examples can be found in Figures~S1 and~S6 of the Text S1). Measured values (solid lines) and extrapolated values are plotted up to the middle of the 21st century (dashed lines). The gray uncertainty range is obtained by including the statistical errors of the regressions (one Standard Deviation (SD) each). For the set of countries for which data is available we obtain the parameters~$h_t$ and~$g_t$ as displayed in the panels~(c) and~(d) of Figure~\ref{fig:examples} for the past values (filled symbols) and for projected values (open symbols). The parameters imply that in average, for a given HDI, the corresponding CO$_2$ emissions decrease during the time frame under investigation, as can also be seen in Figure~S5 of the Text S1. It is apparent that these correlations are hard to overcome since they are intrinsic to the energy supply systems. 

Future country-based emissions estimates are obtained via multiplying the extrapolated CO$_2$ per capita values by population numbers extracted from three scenarios published in the Millennium Ecosystem Assessment report \cite{Alcamo2005}. For the purpose of this work we only make use of data until $2050$ and the population scenarios Adaptive Mosaic~(AM), Technogarden~(TG), and Global Orchestration~(GO). 

The statistical approach undertaken in this work can be named "Development As Usual"~(DAU) in the sense that development and emission trends continue as in the past. Accordingly, we are not claiming that the calculated HDI and CO$_2$ extrapolations are predictions, instead, they represent a plausible near-future world (by 2050) where CO$_2$ emissions from fossil fuel combustion are still closely linked to human development. This assumption is supported by (i) the findings that no discernible decarbonizing trends of energy supply among world regions can be identified \cite{Raupach2007} and (ii) the existence of substantial obstacles to large scale implementation of renewable energy in the near future \cite{Hoffert2002}.

\section*{Results}

\subsection*{Emissions for development}

\begin{figure}
\includegraphics[width=\textwidth]{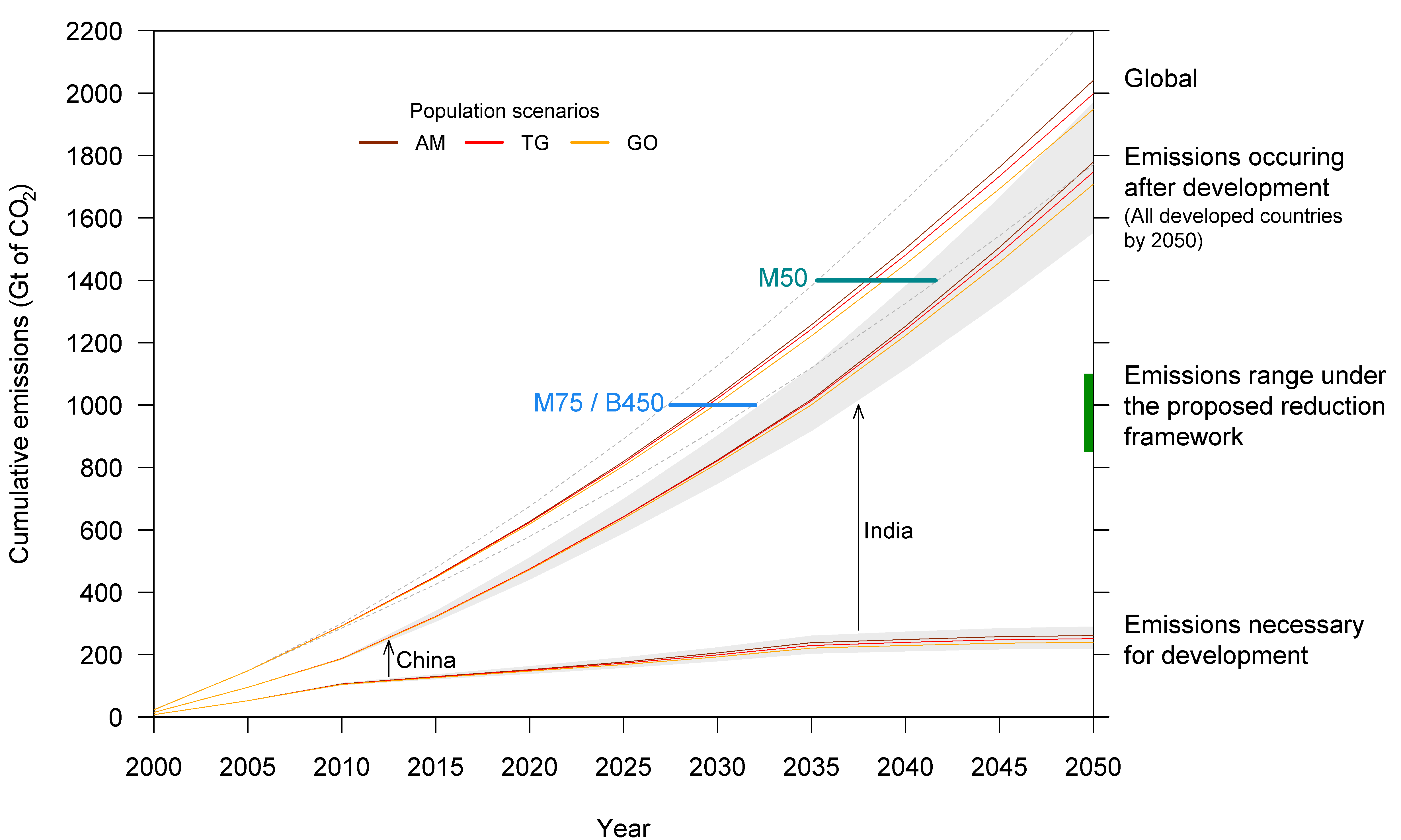}
\caption{
{\bf Cumulative CO$_2$ emissions for Development As Usual (DAU) according to
three population scenarios.} 
Global emissions are split into two emission budgets: emissions necessary for
development (until an HDI of 0.8 is reached) and emissions occurring after
development (all developed countries in 2050).  Population scenarios are
extracted from the Millennium Ecosystem Assessment report \cite{Alcamo2005} (AM
-- Adaptive Mosaic; TG -- Technogarden; GO -- Global Orchestration). Horizontal
lines illustrate the representative values of cumulative CO$_2$ emissions
associated with the probabilities 75\% and 50\% (M75 and M50) of staying below
the $2\,^{\circ}\mathrm{C}$ target by 2050 as provided by Meinshausen \textit{et
al.} \cite{Meinshausen2009} and cumulative emission budgets required to
stabilize CO$_2$ concentrations at 450\,ppm provided by Broecker
\cite{Broecker2007} (B450). The black arrows represent two illustrative examples
(China and India) and indicate the estimated time frame when the HDI threshold
of 0.8 is crossed and emissions are no longer accounted to be necessary for
development. The green bar at the right edge of the frame depicts the range of
cumulative emissions achievable under the proposed reduction framework.
}
\label{fig:projections}
\end{figure}

Figure~\ref{fig:projections} depicts the estimated cumulative emissions for the three population scenarios together with a set of CO$_2$ budgets for particular warming and concentration targets \cite{Broecker2007, Meinshausen2009, Wigley2007}. According to the DAU approach, global cumulative CO$_2$ emissions by 2050 range from~1700 up to~2300\,Gt of CO$_2$ with about 85\% of the world's population living in countries with an HDI above 0.8. When assessed on a per year basis, emissions range between~45.6 and~62.4\,Gt CO$_2$ in 2050 (corresponding respectively to $12.5$ and $17.1$\,Gt of carbon in 2050, using factor 44/12 for conversion \cite{Bowerman2011}), which is within the range of recent projections using IPCC emissions scenarios \cite{Manning2010, Nakicenovic2000}. 

Of a total of 165~countries, 104 were found to be developing countries (HDI
below 0.8) in the year 2000.  By using the UNDP HDI threshold of 0.8 to
differentiate countries with high human development from developing countries
with medium to low human development \cite{UNDP2009}, estimated global CO$_2$ emissions are divided into two budgets. The first budget includes the emissions necessary for the development of countries with HDI below 0.8 while the second budget accounts for emissions occurring after development, this is, emissions from countries with HDI above 0.8. Emissions from countries carrying out a development transition (i.e., crossing the HDI threshold between 2000 and 2050) are added correspondingly to each budget. For example, we estimate India to achieve an HDI above 0.8 between the years 2035 and 2040 (see Table~S1 for the time periods when countries undertake a development transition). Until the HDI threshold is reached the emissions are accounted to be necessary for development, from then on CO$_2$ emissions from India are accounted to occur after development.

In a DAU future we estimate that between~200 and~300\,Gt of cumulative CO$_2$ emissions will be necessary for developing countries (104 in the year 2000) to proceed with their development. In the scope of our approach, 61 developing countries are expected to overcome the HDI of 0.8 by 2050 consuming roughly 98\,\% of the above-mentioned 200-300\,Gt budget. The remaining 43 countries are likely to stay below the UNDP high human development threshold in the considered time frame. Total cumulative emissions occurring after development range from~1500 to~2000\,Gt of CO$_2$.

\begin{table}[h]
\begin{tabular}{ l  c c c }
& \multicolumn{3}{c}{Cumulative CO$_2$ emissions}\\
& \multicolumn{3}{c}{in Gt of CO$_2$ by 2050} \\
\hline
Necessary for development*& {\bf 200} &\hspace{0.5cm}{\bf-} & {\bf 300} \\
\hline
Emitted after development & {\bf 1500} &\hspace{0.5cm}{\bf-} & {\bf 2000}\\
\hspace{0.2cm} {from countries crossing 0.8 HDI between 2000 and 2050} & {\scriptsize 700}  &\hspace{0.5cm}-&  {\scriptsize 1000}\\
\hspace{0.2cm} {from countries already developed in 2000} & {\scriptsize 800}  &\hspace{0.5cm}-& {\scriptsize 1000}\\
\hline
Global &&\hspace{0.5cm}& \\
\hspace{0.2cm}
Emissions under DAU&{\bf 1700}&\hspace{0.5cm}-&{\bf 2300} \\
\hspace{0.2cm}
Emissions under the proposed framework** &{\bf 850}&\hspace{0.5cm}-&{\bf 1100} \\
\hline
\hline
&\multicolumn{3}{}{}\\
&\multicolumn{3}{c}{Allowable CO$_2$ emissions}\\
& \multicolumn{3}{c}{in Gt of CO$_2$}\\
\hline
By 2050\\
\hspace{0.2cm}75\% probability of not exceeding $2\,^{\circ}\mathrm{C}$ \cite{Meinshausen2009} (M75) & \hspace{1.5cm} & {\bf 1000} & \hspace{0.5cm} \\
\hspace{0.2cm}50\% probability of not exceeding $2\,^{\circ}\mathrm{C}$ \cite{Meinshausen2009} (M50) & \hspace{1.5cm} & {\bf 1400} & \hspace{0.5cm} \\
By 2075\\
\hspace{0.2cm}To limit {CO}$_2$ concentrations at 450\,ppm \cite{Broecker2007} (B450) & \hspace{1.5cm} & {\bf 1000} & \hspace{0.5cm} \\ 
\hspace{0.2cm}To limit {CO}$_2$ concentrations at 560\,ppm \cite{Broecker2007} & \hspace{1.5cm} & {\bf 2600} & \hspace{0.5cm} \\
\hspace{0.2cm}To limit {CO}$_2$ concentrations at 560\,ppm \cite{Wigley2007} & \hspace{1.5cm} & {\bf 3300} & \hspace{0.5cm} \\
\hline
\hline
\end{tabular}
\caption{
{\bf Projected cumulative CO$_2$ emissions for the period 2000-2050 compared to CO$_2$ emission budgets for warming potential and atmospheric concentrations.}
The table summarizes the emission values before and after countries reach the
HDI of 0.8 according to a DAU approach and under the proposed reduction
framework. A collection of previous calculated budgets for allowable CO$_2$
emissions highlights the efforts necessary for emission reductions. * Cumulative
emissions necessary for development assuming an HDI threshold of 0.9 would range
from 700 to 900\,Gt CO$_2$. **Assuming the same uncertainty as in DAU.}
\label{tab:numbers}
\end{table}

This amount is similarly divided among countries carrying out a development
transition (700 to~1000\,Gt) and those whose development occurred before the
year 2000 (800 to~1000\,Gt) as summarized in Table~\ref{tab:numbers}.

Emissions for development where found to be very sensitive to the selected HDI score. Assuming that developing countries want to achieve western development styles would require to set the minimum development standards to values of 0.9. In such a case, emissions necessary for development by 2050 range from about 700 to 900\,Gt of CO$_2$. This range is higher by at least a factor of 3 than the values obtained for a HDI threshold of 0.8.

We further compare our estimates with previously calculated CO$_2$ budgets for particular time frames, global warming targets and atmospheric CO$_2$ concentrations. We find that the emissions necessary for development consume up to 30\,\% of the 1000\,Gt CO$_2$ limit for a 75\,\%~probability of keeping global warming below $2\,^{\circ}\mathrm{C}$, as calculated by Meinshausen \textit{et al.} \cite{Meinshausen2009} and indicated as M75 in Fig.~\ref{fig:projections}. According to our projections, the 1000\,Gt budget limit by 2050 would already be exhausted around 2030 if human development proceeds as in the past. In case one adopts the CO$_2$ limit providing a 50\,\% chance (M50) of staying below $2\,^{\circ}\mathrm{C}$, then cumulative CO$_2$ emissions necessary for development would still represent about 20\,\% of the total budget. Similarly, the CO$_2$ budget to stabilize atmospheric concentrations at 450\,ppm provided by Broecker \cite{Broecker2007} (indicated as B450 in Fig.~\ref{fig:projections}), would be exhausted within the next 20~years. 

\subsection*{Human development framework for CO$_2$ allocation and reduction}

\begin{figure}
\includegraphics[width=\textwidth]{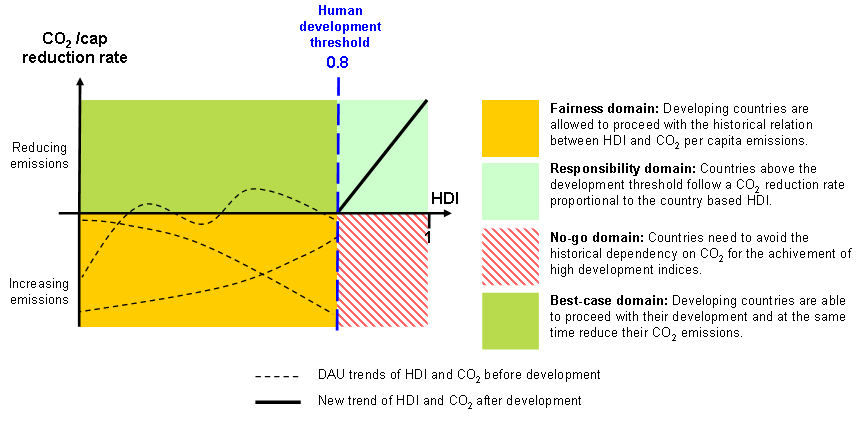}
\caption{
{\bf CO$_2$ emissions reduction framework based on HDI.}
The reduction framework proposes four domains of climate action that are both
fair in an historical perspective and constrained by current technological
developments. Reserving a fairness domain for developing countries implies that
their participation in climate efforts can be operated in a voluntary basis. The
development threshold of 0.8 HDI is taken from United Nations Development Report
2009 \cite{Bongaarts2010}.
}
\label{fig:frame}
\end{figure}

The question logically arising from the results is how to operate a fair
transition of developing countries towards high development standards without
compromising current climate targets. A fair approach implies that an
hypothetical developing country should not be limited in its emissions of
CO$_2$ until it reaches a particular threshold of human development. In practice, the development path made by current developed countries in the past should be possible for developing countries in the future if they choose to do so. This key aspect of the proposed framework convenes in our opinion a better representation of fairness in CO$_2$ emissions allocation as opposed to fixing a point in the past from where emissions are integrated. Figure~\ref{fig:frame} makes use of the 0.8 HDI threshold to differentiate four areas of action regarding climate policies. Countries whose HDI trails below the minimum human development standard evolve in the context of a \textit{Fairness domain}. In this domain the developing country is allowed to fulfill the basic development needs by following a development path where HDI is highly correlated with CO$_2$ emissions from fossil fuel burning. In the \textit{Best-case domain} developing countries are able to proceed with their development goals and at the same time reduce their CO$_2$ emissions. This domain would imply a fast worldwide implementation of energy technologies with low carbon intensity, a transformation that is not observed so far \cite{Raupach2007}. After basic development needs are fulfilled, countries are no longer said to be developing and transit to the \textit{Responsibility domain} where they engage in CO$_2$ reduction rates proportional to their HDI in order to preserve a global warming limit of $2\,^{\circ}\mathrm{C}$ by 2050 \cite{Meinshausen2009}. The \textit{No-go domain} needs to be avoided by future developed countries and quickly abandoned by current ones on the basis that resulting emissions would be largely incompatible with future climatic policies. A generalized convergence of countries towards the \textit{Responsibility domain} should be operated. 

\begin{figure}
\includegraphics[width=0.75\textwidth]{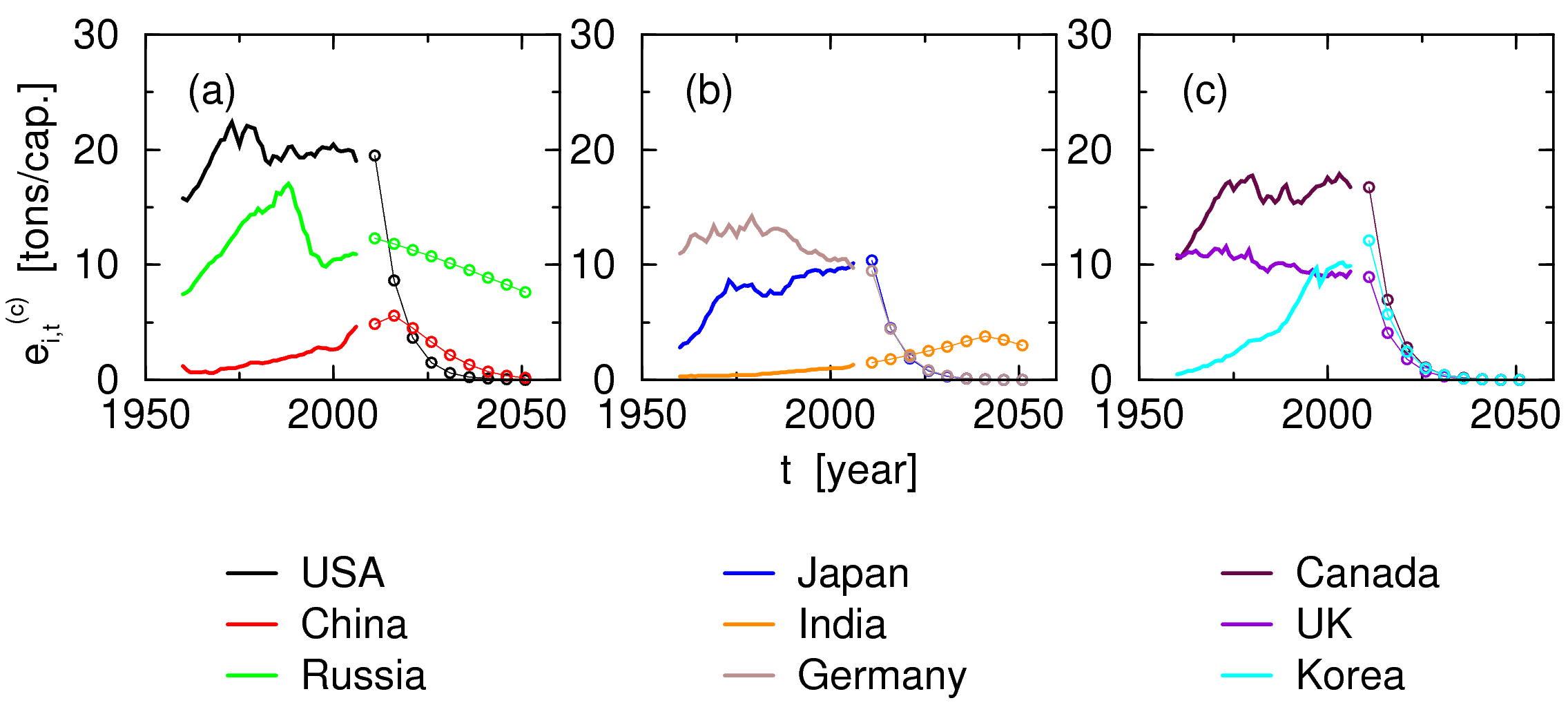}
\caption{
{\bf Examples of extrapolated CO$_2$ emissions per capita in agreement with the
proposed reduction scheme ($d^{*}=0.8$, $f=3.3$).}
Solid lines stand for the historical emission while the connected circles
represent extrapolated emissions when countries follow the reduction scheme
proposed.
}
\label{fig:reduction_exp}
\end{figure}
 
To formalize this, we propose that a developed country~$i$ reduces it's per capita emissions 
at year~$t$ according to 
$e^{\rm (c)}_{i,t-5\text{y}}\rightarrow(1- r_{i,t})\,e^{\rm (c)}_{i,t}$
with the 5-year reduction rate  $r_{i,t}$, given by
\begin{equation}
\label{eq:co2redux}
r_{i,t} = f\,(d_{i,t}-d^{*}) \quad \text{for} \quad d_{i,t}>d^{*}
\end{equation}
where $d^{*}$ is the development threshold and $f$ a proportionality constant 
which determines how strong the reduction rate increases with increasing HDI
(see also Text S1 section~V). Based on the above discussed development
threshold ($d^{*}=0.8$) we estimate that $f=3.3$ (as a lower bound) would lead to global cumulative emissions ranging between 850 and 1100\,Gt of CO$_2$ by 2050 if reduction starts in 2015 (assuming the same uncertainty as in DAU). This amount is within the range of allowed cumulative CO$_2$ emissions that provide between 80\,\% and 66\,\% change of keeping global temperatures below a $2\,^{\circ}\mathrm{C}$ increase, as calculated by \cite{Meinshausen2009}. Under our reduction framework, global emissions in the year 2050 are estimated to be 10\,Gt CO$_2$ or about 13.3\,Gt CO$_2$ equivalent if one accounts also for non-CO$_2$ gases (with non-CO2 gases constituting roughly 1/3 of total CO$_2$ equivalent \cite{Meinshausen2009}). This value is relatively low and complies with post-2050 emission thresholds that make cumulative CO$_2$ emissions between 2010 and 2050 a robust indicator of achieving the $2\,^{\circ}\mathrm{C}$ target as in Meinshausen \textit{et al.} \cite{Meinshausen2009} and Bowerman \textit{et al.} \cite{Bowerman2011}.

\begin{figure}
\includegraphics[width=\textwidth]{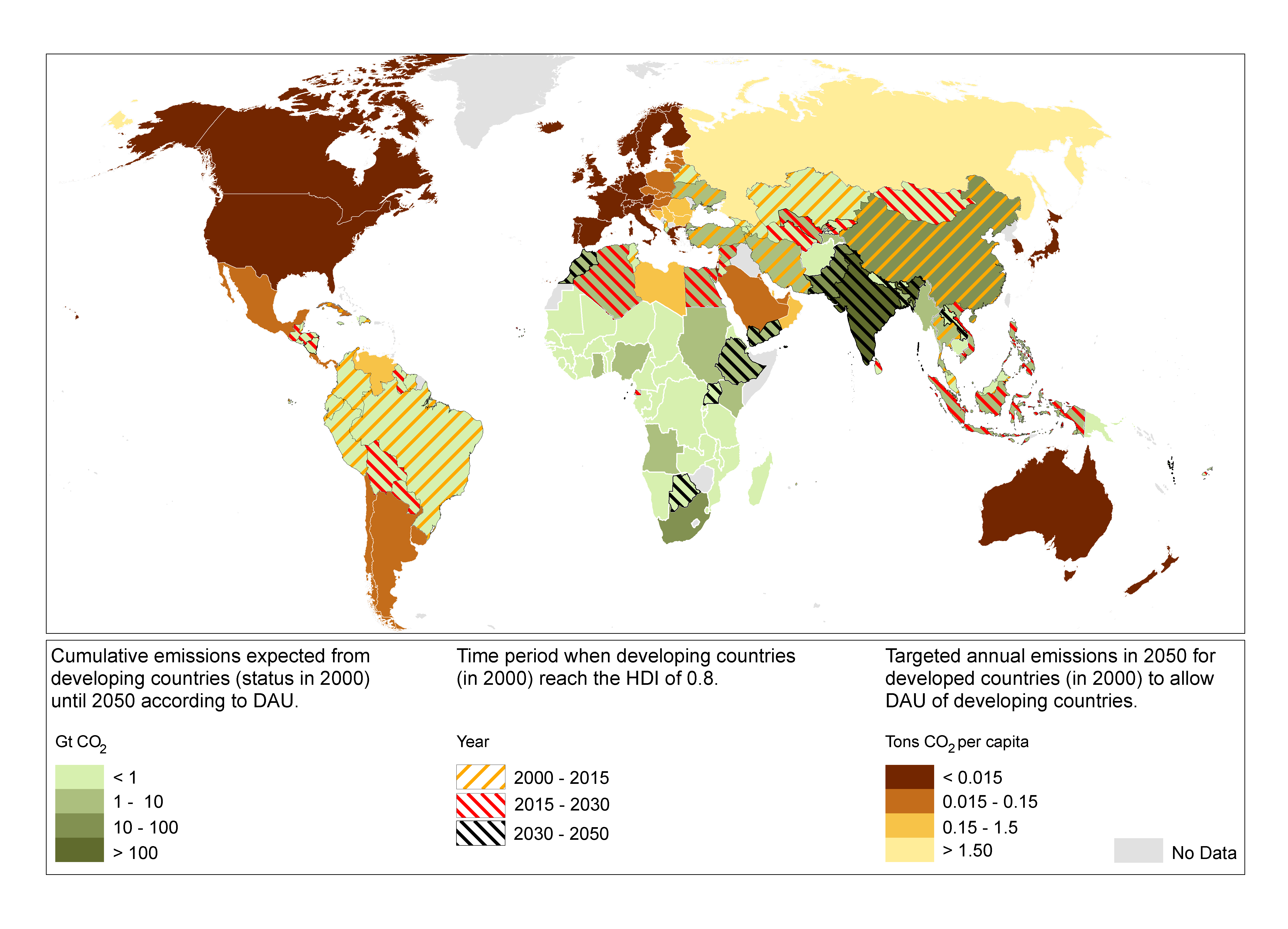}
\caption{
{\bf Global distribution of allowed emissions for DAU from developing countries
(green shading) and per capita CO$_2$ targets in 2050 for developed countries
(brown shading) under the proposed framework to keep temperatures below
$2\,^{\circ}\mathrm{C}$ target -- as implied by the M75 CO$_2$ budget.} 
The period in time when developing countries are expected to reach an HDI of 0.8
is represented by the colored hatches. 
}
\label{fig:reduction_map}
\end{figure}

The value of $f=3.3$ implies that in each time step of five years, countries with an HDI of $0.85$ would need to reduce their per capita emissions by approx. $17$\,\% and countries with an HDI of $0.9$ by $33$\,\%. As a result of applying these reduction rates, emission curves of current and future developed countries decrease approximately exponentially. In Figure~\ref{fig:reduction_exp} we show the emission trajectories for a set of countries. Per capita CO$_2$ emissions from Germany would need to be reduced from about $10$ tons in 2010 to $4$ tons in 2020 and a nearly complete decarbonization by 2040. Countries not yet developed are entitled to increase emissions. In the case of India, CO$_2$ emissions  per capita grow until a maximum of $4$ tons in 2040. After its development, India needs to reduce per capita emissions to approx.~3.5 and 3~tons CO$_2$ in 2045 and 2050 respectively. Developing countries unable to reach an HDI of 0.8 during the time frame of this analysis are allowed to emit following DAU. For example, Pakistan is entitled to increase emissions to a maximum of approx.~2.5 tons per capita in 2050, the year when its expected to become a developed country following our approach. In Figure~\ref{fig:reduction_map} we provide an overview of our results according to the current political world map. The figure highlights the geographic trade-offs between the necessary achievements in CO$_2$ reduction by current developed countries (brown shading), and the cumulative CO$_2$ emissions for the DAU of developing countries (green shading) in order to comply with the $2\,^{\circ}\mathrm{C}$ target -- using the M75 budget. 

\section*{Discussion}

Previous reduction schemes of global CO$_2$ emissions make use of
population numbers \cite{WBGU2006,Broecker2007,Meinshausen2007} or income
distribution \cite{Baer2008} associated with permissible CO$_2$ atmospheric
concentrations or global warming targets \cite{Zickfeld2009} to comply with the
"common but shared responsibility" principle of the 1992 United Nations
Framework Convention on Climate Change (UNFCCC). 

These approaches disregard to some extent the possible development set-backs caused by CO$_2$ reductions in the socio-economic development of a country. We use the HDI in order to take development needs of developing countries into consideration. In a DAU world, we estimate that up to 300\,Gt of CO$_2$ represent a pre-condition for raising a considerable amount of developing countries (comprising HDI below 0.8 in the year 2000) to a minimum HDI of 0.8 in the year 2050. If development pathways proceed as in the past, resulting CO$_2$ emissions will pose tighter constraints on the achievement of the previously mentioned climate targets. One can legitimately question the likelihood of such assumption. In a sense, our approach can only be regarded as an approximation since aspects like technological innovation and enhanced technology transfer between developed and developing countries cannot be anticipated. This is a recurrent problem when projecting trends of socio-economic systems into the future. The assumed ergodicity would benefit from further investigation.

Depending on mankind's decision concerning acceptable levels of climate change and desirable human development goals, emissions necessary for development can represent substantial shares of the CO$_2$ budgets here analyzed (see Table~\ref{tab:numbers} for further CO$_2$ budgets). In line with previous research \cite{Steinberger2010}, it was found that the overall efficiency in achieving higher human development scores increases, e.g. less CO$_2$ emissions are necessary for a certain HDI. It remains open to which extent these gains in efficiency can be articulated in the context of current climate negotiations constraints. 

We propose a differentiated and dynamic allocation scheme of CO$_2$ emissions based on human development achievements. Developing countries are not obliged to reduce their emissions until a certain threshold of human development is achieved. From then on the country is no longer considered to be developing, and should therefore engage on the proposed emissions reduction path. It is worth to point out that the investigated population scenarios only show substantial divergence in values beyond 2050. Obtained differences in CO$_2$ emissions between scenarios are therefore small during the time frame of analysis.

Within the scope of our approach the efforts for climate protection commitments from developing countries can be operated on a voluntary basis. With CO$_2$ reduction rates linked to the evolution of HDI as proposed here, the $2\,^{\circ}\mathrm{C}$ target can be met even if emissions from developing countries evolve according to DAU during the early stages of development. Independent of the climate target, a fair allocation and reduction of emissions between developed and developing countries must consider the dependence between CO$_2$ and human development here discussed.

\section*{Acknowledgments}

The authors acknowledge the financial support from BaltCICA (Baltic Sea Region
Programme 2007-2013). They wish to thank the Federal Ministry for the
Environment, Nature Conservation, and Nuclear Safety of Germany who supports
this work within the framework of the International Climate Protection
Initiative.
We thank M. Boettle, S. Havlin, A. Holsten, D. Reusser, H.D. Rozenfeld, J.
Sehring, and J. Werg for discussions and comments. Furthermore, we thank A.
Schlums for help with the manuscript. The main author further thanks N.
Kozhevnikova for her refined sense of critique. All authors express their
gratitude for the Editor comments that largely benefited the current manuscript.


\begin{thebibliography}{10}
\providecommand{\url}[1]{\texttt{#1}}
\providecommand{\urlprefix}{URL }
\expandafter\ifx\csname urlstyle\endcsname\relax
  \providecommand{\doi}[1]{doi:\discretionary{}{}{}#1}\else
  \providecommand{\doi}{doi:\discretionary{}{}{}\begingroup
  \urlstyle{rm}\Url}\fi
\providecommand{\bibAnnoteFile}[1]{%
  \IfFileExists{#1}{\begin{quotation}\noindent\textsc{Key:} #1\\
  \textsc{Annotation:}\ \input{#1}\end{quotation}}{}}
\providecommand{\bibAnnote}[2]{%
  \begin{quotation}\noindent\textsc{Key:} #1\\
  \textsc{Annotation:}\ #2\end{quotation}}
\providecommand{\eprint}[2][]{\url{#2}}

\bibitem{Elzen2008}
den Elzen M, H\"ohne N (2008) Reductions of greenhouse gas emissions in
  annex~{I} and non-annex~{I} countries for meeting concentration stabilisation
  targets.
\newblock Clim Change 91: 249 - 274.
\input{Elzen2008}

\bibitem{Rogelj2010}
Rogelj J, Nabel J, Chen C, Hare W, Markmann K, et~al. (2010) Copenhagen accord
  pledges are paltry.
\newblock Nature 464: 1126 - 1128.
\input{Rogelj2010}

\bibitem{Cocklin2007}
Cocklin C, Heller T, Lecocq F, Llanes-Regueiro J, Pan J, et~al. (2007)
  Mitigation. Contribution of Working Group III to the Fourth Assessment Report
  of the Intergovernmental Panel on Climate Change.
\newblock Cambridge University Press, Cambridge.
\input{Cocklin2007}

\bibitem{Bolin2001}
Bolin B, Kheshgi HS (2001) On strategies for reducing greenhouse gas emissions.
\newblock Proc Nat Acad Sci USA 98: 4850 - 4854.
\input{Bolin2001}

\bibitem{Broecker2007}
Broecker WS (2007) {CO}$_2$ arithmetic.
\newblock Science 315: 1371.
\input{Broecker2007}

\bibitem{Chakravarty2010}
Chakravarty S, Chikkatur A, de~Coninck H, Pacala S, Socolow R, et~al. (2010)
  Sharing global {CO}$_2$ emission reductions among one billion high emitters.
\newblock Proc Nat Acad Sci USA 106: 11884 - 11888.
\input{Chakravarty2010}

\bibitem{WBGU2006}
WBGU (2006) Solving the climate dilemma: The budget approach.
\newblock German Advisory Council on Global Change.
\input{WBGU2006}

\bibitem{IPCC2001}
IPCC (2001) Climate Change 2001: Synthesis Report. A Contribution of Working
  Groups I, II, and III to the Third Assessment Report of the Integovernmental
  Panel on Climate Change.
\newblock Cambridge University Press.
\input{IPCC2001}

\bibitem{Pan2005}
Jiahua P (2005) Meeting human development goals with low emissions: An
  alternative to emissions caps for post-kyoto from a developing country
  perspective.
\newblock International Environmental Agreements: Politics, Law and Economics
  5: 89-104.
\input{Pan2005}

\bibitem{Baer2008}
Baer P, Athanasiou T, Kartha S, Kemp-Benedict E (2008) The Greenhouse
  Development Rights Framework: The right to development in a climate
  constrained world (Revised second edition).
\newblock Heinrich Boell Foundation and Christian Aid and EcoEquity and
  Stockholm Environment Institute.
\input{Baer2008}

\bibitem{UNDP2008}
UNDP (2008) Human development report, techical note 1: Calculating the human
  development indices.
\newblock Technical report, United Nations.
\input{UNDP2008}

\bibitem{Neumayer2001}
Neumayer E (2001) The human development index and sustainability--a
  constructive proposal.
\newblock Ecol Econ 39: 101 - 114.
\input{Neumayer2001}

\bibitem{Atkinson1997}
Atkinson G, Dubourg R, Hamilton K, Munasinghe M, Pearce D, et~al. (1997)
  Measuring Sustainable Development - Macroeconomics and the Environment.
\newblock Edward Elgar Publishing.
\input{Atkinson1997}

\bibitem{Patt2010}
Patt A, Tadross M, Nussbaumer P, Asante K, Metzger M, et~al. (2010) Estimating
  least-developed countries' vulnerability to climate-related extreme events
  over the next 50 years.
\newblock Proc Nat Acad Sci USA doi: 10.1073/pnas.0910253107: 1 - 5.
\input{Patt2010}

\bibitem{Brooks2005}
Brooks N, Adger W, Kelly P ({2005}) {The determinants of vulnerability and
  adaptive capacity at the national level and the implications for adaptation}.
\newblock Global Environ Chang {15}: {151-163}.
\input{Brooks2005}

\bibitem{Montalvo2010}
Montalvo JG, Ravallion M ({2010}) {The pattern of growth and poverty reduction
  in China}.
\newblock J Compar Econ {38}: {2-16}.
\input{Montalvo2010}

\bibitem{Schellnhuber2006}
Schellnhuber HJ, Cramer W, Nakicenovic N, Wigley T, Yohe G (2006) Avoiding
  Dangerous Climate Change, Chapter 29.
\newblock Cambridge University Press, 281 pp.
\input{Schellnhuber2006}

\bibitem{HosmerL2000}
Hosmer DW, Lemeshow S (2000) Applied Logistic Regression.
\newblock Wiley Series in Probability and Statistics. New York: John Wiley \&
  Sons.
\input{HosmerL2000}

\bibitem{MalmgrenSCA2009}
Malmgren RD, Stouffer DB, Campanharo ASLO, Amaral LAN (2009) On universality in
  human correspondence activity.
\newblock Science 325: 1696--1700.
\input{MalmgrenSCA2009}

\bibitem{Rhemtulla2008}
Rhemtulla JM, Mladenoff DJ, Clayton MK (2008) Historical forest baselines
  reveal potential for continued carbon sequestration.
\newblock Proc Nat Acad Sci USA 106: 6082 - 6087.
\input{Rhemtulla2008}

\bibitem{MasonGH1989}
Mason RL, Gunst RF, Hess JL (1989) Statistical Design and Analysis of
  Experiments -- with applications to engineering and science.
\newblock New York: John Wiley \& Sons.
\input{MasonGH1989}

\bibitem{Alcamo2005}
Alcamo J, Alder J, Bennett E, Carr ER, Deane D, et~al. (2005) Millennium
  Ecosystem Assessment.
\newblock Island Press.
\input{Alcamo2005}

\bibitem{Raupach2007}
Raupach MR, Marland G, Ciais P, Quéré CL, Canadell JG, et~al. (2007) Global and
  regional drivers of accelerating {CO}$_2$ emissions.
\newblock Proc Nat Acad Sci USA 104: 10288 - 10293.
\input{Raupach2007}

\bibitem{Hoffert2002}
Hoffert M, Caldeira K, Benford G, Criswell D, Green C, et~al. (2002) Advanced
  technology paths to global climate stability: Energy for a greenhouse planet.
\newblock Science 298: 981 - 987.
\input{Hoffert2002}

\bibitem{Meinshausen2009}
Meinshausen M, Meinshausen N, Hare W, Raper S, Frieler K, et~al. (2009)
  Greenhouse-gas emission targets for limiting global warming to
  $2\,^{\circ}\mathrm{C}$.
\newblock Nature 458: 1158 - 1162.
\input{Meinshausen2009}

\bibitem{Wigley2007}
Wigley T (2007) {CO}$_2$ emissions: A piece of the pie.
\newblock Science 316: 829 - 830.
\input{Wigley2007}

\bibitem{Bowerman2011}
Bowerman NHA, Frame DJ, Huntingford C, Lowe JA, Allen MR (2011) Cumulative
  carbon emissions, emissions floors and short-term rates of warming:
  implications for policy.
\newblock Phil Trans R Soc A 369: 45-66.
\input{Bowerman2011}

\bibitem{Manning2010}
Manning MR, Edmonds J, Emori S, Grubler A, Hibbard K, et~al. (2010)
  Misrepresentation of the {IPCC} {CO}$_2$ emission scenarios.
\newblock Nature Geosci 3: 376 - 377.
\input{Manning2010}

\bibitem{Nakicenovic2000}
Nakicenovic N, Swart R (2000) IPCC Special Report on Emission Scenarios.
\newblock Cambridge University Press.
\input{Nakicenovic2000}

\bibitem{UNDP2009}
UNDP (2009) Human Development Report 2009: Overcoming barriers: Human mobility
  and development.
\newblock Palgrave Macmillan.
\input{UNDP2009}

\bibitem{Meinshausen2007}
Meinshausen M (2007) Stylized emission path.
\newblock Background note on stylized emission pathway for UNDP Human
  Development Report 2007 .
\input{Meinshausen2007}

\bibitem{Zickfeld2009}
Zickfeld K, Eby M, Matthews HD, Weaver AJ (2009) Setting cumulative emissions
  targets to reduce the risk of dangerous climate change.
\newblock Proc Nat Acad Sci USA 106: 16129 - 16134.
\input{Zickfeld2009}

\bibitem{Steinberger2010}
Steinberger JK, Roberts JT ({2010}) {From constraint to sufficiency: The
  decoupling of energy and carbon from human needs, 1975-2005}.
\newblock Ecol Econ {70}: {425-433}.
\input{Steinberger2010}

\bibitem{Bongaarts2010}
Bongaarts J, Greenhalgh S, McNicoll G ({2010}) {Human Development Report 2009:
  Overcoming Barriers: Human Mobility and Development}.
\newblock Population Devel Rev {36}: {402-403}.
\input{Bongaarts2010}

\end{thebibliography}

\end{document}